\documentclass[journal=jctcce,manuscript=article]{achemso}
\setkeys{acs}{usetitle=true}

\usepackage[version=3]{mhchem} 

\usepackage{amsmath,amssymb}
\usepackage{graphicx,color}
\usepackage{amsfonts}
\usepackage{algorithm,algorithmic}
\usepackage{epstopdf}
\usepackage{comment}
\usepackage{subfig}

\usepackage[version=3]{mhchem} 

\newcommand{\Or}{\mathcal{O}}

\newcommand{\abs}[1]{\left\lvert#1\right\rvert}
\newcommand{\norm}[1]{\left\lVert#1\right\rVert}

\newcommand{\ud}{\,\mathrm{d}}

\newcommand{\wt}[1]{\widetilde{#1}}

\newcommand{\bvec}[1]{\mathbf{#1}}
\newcommand{\vr}{\bvec{r}}

\newcommand{\I}{\imath} 

\newcommand{\vace}{\widetilde{V}_{\mathrm{X}}}

\newcommand{\ext}{\mathrm{ext}}
\newcommand{\Hxc}{\mathrm{Hxc}}
\newcommand{\X}{\mathrm{X}}

\newcommand{\REV}[1]{{#1}}
\newcommand{\angstrom}{\mbox{\normalfont\AA}~}

\title[Fast Hybrid rt-TDDFT]{Fast real-time time-dependent hybrid functional calculations with 
the parallel transport gauge and the adaptively compressed exchange formulation}

\author{Weile Jia}
\affiliation{Department of Mathematics, University of California, 
Berkeley, California 94720, United States}

\author{Lin Lin}
\email{linlin@math.berkeley.edu}
\affiliation{Department of Mathematics, University of California, 
Berkeley, California 94720, United States}
\affiliation{Computational Research Division, Lawrence Berkeley 
National Laboratory, Berkeley, California 94720, United States}

\begin{document}
\begin{abstract}

\begin{figure}[h]
  \begin{center}
    \includegraphics[width=0.40\textwidth]{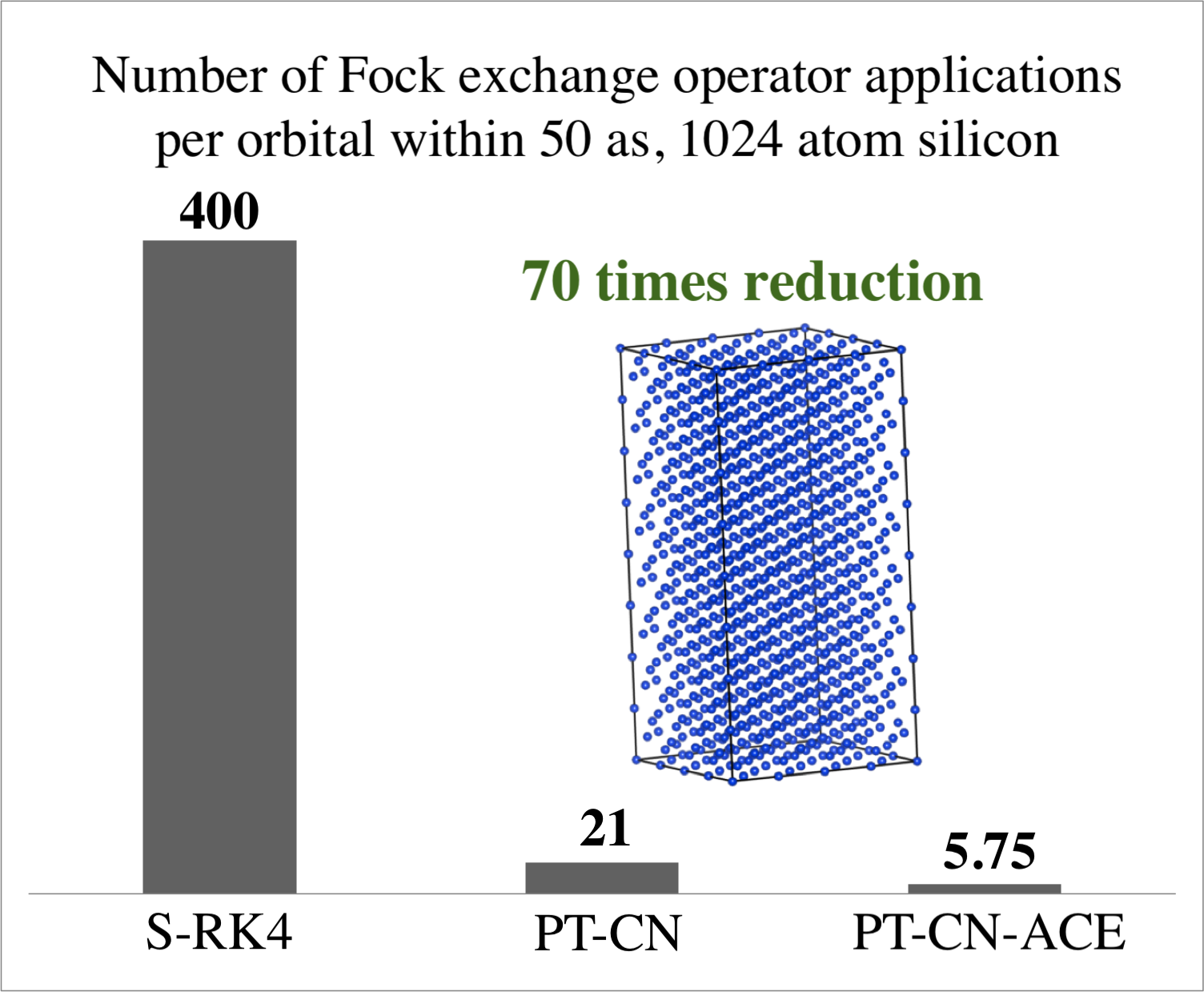}
  \end{center}
  \label{fig:si1024-phi}
\end{figure}

We present a new method to accelerate real time-time dependent density
functional theory (rt-TDDFT) calculations with hybrid
exchange-correlation functionals. In the context of a large basis set
such as planewaves and real space grids, the main computational
bottleneck for large scale calculations is the application of the Fock
exchange operator to the time-dependent orbitals.
Our main goal is to reduce the frequency of applying the Fock exchange
operator, without loss of accuracy.  We achieve this by combining the
recently developed parallel transport (PT) gauge formalism [Jia et al,
arXiv:1805.10575] and the adaptively compressed exchange operator (ACE) formalism
[Lin,  J. Chem. Theory Comput. 12, 2242, 2016].  The PT gauge yields the slowest possible dynamics among all choices of gauge. When coupled with implicit time integrators such as
the Crank-Nicolson (CN) scheme, the resulting PT-CN scheme can
significantly increase the time step from sub-attoseconds to $10-100$
attoseconds.  At each time step $t_{n}$, PT-CN requires the
self-consistent solution of the orbitals at time $t_{n+1}$. We use
ACE to delay the update of the Fock exchange operator in this nonlinear
system, while maintaining the same self-consistent solution. We verify the performance of the
resulting PT-CN-ACE method by computing the absorption spectrum of a
benzene molecule and the response of bulk silicon systems to an
ultrafast laser pulse, using the planewave basis set and the HSE
exchange-correlation functional. We report the strong and weak scaling
of the PT-CN-ACE method for silicon systems ranging from $32$ to $1024$
atoms, on a parallel computer with up to $2048$ computational cores. Compared to
standard explicit time integrators such as the $4th$ order Runge-Kutta
method (RK4), we find that the PT-CN-ACE can reduce the frequency of the
Fock exchange operator application by nearly $70$ times, and the
thus reduce the overall wall clock time time by 46 times for the system
with $1024$ atoms. Hence our work enables hybrid functional rt-TDDFT
calculations to be routinely performed with a large basis set for the
first time.  
\end{abstract}

\section{Introduction} \label{sec:Introduction}

In Kohn-Sham density functional
theory~\cite{HohenbergKohn1964,KohnSham1965}, hybrid exchange-correlation functionals, such as
B3LYP~\cite{LeeYangParr1988,Becke1993},
PBE0~\cite{PerdewErnzerhofBurke1996} and
HSE~\cite{HeydScuseriaErnzerhof2003,HeydScuseriaErnzerhof2006}, are known to be
more reliable in producing high fidelity results for ground state electronic
structure calculations \REV{for a vast range of systems~\cite{AStroppaGkresse2008, MMarsmanetal2008}}.  With the recent developments of ultrafast
laser techniques, a large number of excited state phenomena, such as nonlinear
optical response~\cite{TakimotoVilaRehr2007} and the collision of an ion
with a substrate~\cite{KrasheninnikovMiyamotoTomanek2007}, can be
observed in real time.  One of the most widely used techniques for
studying such ultrafast properties is the real-time time-dependent
density functional theory
(rt-TDDFT)~\cite{RungeGross1984,YabanaBertsch1996,OnidaReiningRubio2002,AndreaniColognesiMayersEtAl2005,AlonsoAndradeEcheniqueEtAl2008,Ullrich2011}. Hybrid
functional rt-TDDFT calculations have been performed in the context of
small basis sets such as Gaussian orbitals and atomic
orbitals~\cite{LopataGovind2011,LopataVanKhalilEtAl2012,DingVanEichingerEtAl2013}.
In the context of a large basis set such as planewaves and real space
grids, to the extent of our knowledge, all rt-TDDFT calculations so far are
performed with local and semi-local exchange-correlation functionals.
This is because hybrid functionals include a fraction of the
Fock exchange operator, which requires access to the diagonal as
well as  the off-diagonal elements of the density matrix.  This leads to
significant increase of the computational cost compared to calculations
with local and semi-local functionals. The problem is compounded by the
very small time step (on the order of attosecond or sub-attosecond)
often needed in rt-TDDFT simulation. Hence to reach the femtosecond, let
alone the picosecond timescale, the number of application of the Fock
exchange operator can be prohibitively expensive for large systems.  

This paper aims at enabling practical hybrid functional rt-TDDFT
calculations to be performed with a large basis set. To this end, the
primary goal is to reduce the number of matrix-vector multiplication
operations involving the Fock exchange operator. At first glance, this
is a rather difficult task, since the Fock exchange operator depends on
the density matrix $P(t)$, which needs to be updated at each small time
step.  In order to overcome this difficulty, we first enlarge the time
step of rt-TDDFT calculations via implicit time integrators.  While
implicit time integrators are often considered to be not sufficiently
cost-effective when compared to explicit integrators for rt-TDDFT
calculations~\cite{CastroMarquesRubio2004,SchleifeDraegerKanaiEtAl2012,GomezMarquesRubioEtAl2018},
these studies are performed by direct propagation of the Kohn-Sham
wavefunctions. One main reason is that the oscillation of the
wavefunctions is faster than that of density matrix, and hence
implicit integrators with a large time step are often stable but not
accurate enough.
We have recently identified that by optimizing the gauge, i.e. a
unitary rotation matrix performing a linear combination of the
wavefunctions, the oscillation of the wavefunctions can be significantly
reduced. In particular, the parallel transport (PT)
gauge~\cite{AnLinPT,JiaAnWangEtAlPT} yields the {\emph slowest 
dynamics} among all possible choices of gauge.  Hence when the parallel
transport dynamics is coupled with implicit time integrators, such as the
Crank-Nicolson (CN) scheme, the resulting PT-CN scheme can significantly
increase the time step with systematically controlled accuracy. 

Implicit integrators such as PT-CN introduces a set of nonlinear
equations that needs to be solved self-consistently at each time step
going from $t_{n}$ to $t_{n+1}$. This system can be viewed as a fixed
point problem to determine the orbitals at $t_{n+1}$. In the context of
ground state hybrid functional density functional theory calculations,
we have recently developed the adaptively compressed exchange operator
(ACE) formulation~\cite{Lin2016ACE,HuLinBanerjeeEtAl2017} to accelerate a fixed point problem
introduced by a nonlinear eigenvalue problem.  The ACE formulation can
reduce the frequency of applying the Fock exchange operator without loss of
accuracy, and can be used for insulators and metals. It has been
incorporated into community software packages such as the Quantum
ESPRESSO~\cite{GiannozziBaroniBoniniEtAl2009}. The idea of the adaptive
compression has been rigorously analyzed in the context of linear
eigenvalue problems~\cite{LinLindseyACE}, and can be extended to
accelerate calculations in other contexts such as the density functional
perturbation theory~\cite{LinXuYing2017}.  In this paper, we further
extend the idea of ACE to accelerate hybrid functional rt-TDDFT
calculations, by splitting the solution of the nonlinear
system into two iteration loops.  During each iteration of the outer
loop, we only apply the Fock exchange operator once per orbital, and
construct the adaptively compressed Fock exchange operator. In the inner loop,
only the adaptively compressed Fock exchange operator will be used, and the
application of the compressed operator only involves matrix-matrix
multiplication operations, and is much cheaper than applying the Fock exchange operator. This
two loop strategy further reduces the frequency of updating the Fock
operator and hence the computational time.

The rest of the manuscript is organized as follows. We introduce the
real-time time-dependent density functional theory with hybrid functional and
parallel transport gauge in section ~\ref{sec:rttddft} and ~\ref{sec:pt}, 
respectively. We present the adaptively compressed exchange operator formulation in
section ~\ref{sec:pcdiis}. Numerical results are presented in
section~\ref{sec:Result}, followed by a conclusion and discussion in
section~\ref{sec:conclusion}.

\section{Real-time time dependent functional theory with hybrid
functional}\label{sec:rttddft}

rt-TDDFT solves the following set of time-dependent equations
\begin{equation} 
  \I \partial_{t} \psi_{i}(t) = H(t,P(t)) \psi_{i}, \quad
  i=1,\ldots,N_{e},
  \label{eqn:tddft}
\end{equation}
where $N_e$ is the number of electrons (spin degeneracy omitted), and
$\Psi(t)=[\psi_{1}(t),\ldots,\psi_{N_{e}}(t)]$ are the electron orbitals.
The Hamiltonian takes the form
\begin{equation}
  H(t,P(t)) = -\frac12 \Delta_{\vr} + V_{\ext}(t) + V_{\Hxc}[P(t)] +
  V_{\X}[P(t)].
  \label{}
\end{equation}
Here $V_{\ext}(t)$ characterizes the electron-ion interaction, and the explicit dependence of the
Hamiltonian on $t$ is often due to the existence of an external field.
The Hamiltonian also depends nonlinearly on the density matrix $P(t) =
\Psi(t)\Psi^{*}(t)$.  $V_{\Hxc}$ is a local operator, and characterizes the Hartree
contribution and the local and the semi-local part of the exchange-correlation contribution. It depends only on the
electron density $\rho(t) = \sum_{i=1}^{N_e}|\psi_i(t)|^2$, which are
given by the diagonal matrix elements of the density matrix $P(t)$ in
the real space representation.
The Fock exchange operator $V_{\X}$ is an integral operator
with kernel
\begin{equation}
  V_{\X}[P](\vr,\vr') = -\alpha P(\vr,\vr') K(\vr,\vr').
  \label{eqn:VXkernel}
\end{equation} 
Here $K(\vr,\vr')$ is the kernel for the electron-electron interaction,
and $0<\alpha<1$ is a mixing fraction.
For example, in the Hartree-Fock theory, $K(\vr,\vr')=1/\abs{\vr-\vr'}$
is the Coulomb operator and $\alpha=1$. In screened exchange
theories~\cite{HeydScuseriaErnzerhof2003}, $K$
can be a screened Coulomb operator with kernel
$K(\vr,\vr')=\text{erfc}(\mu\abs{\vr-\vr'})/\abs{\vr-\vr'}$, and
typically $\alpha\sim 0.25$.

When a large basis set is used, it is
prohibitively expensive to explicitly construct $V_{\X}[P]$, and it is only
viable to apply it to a vector $\varphi(\vr)$ as
\begin{equation}
  \left(V_{X}[P]\varphi\right)(\vr) = 
  -\sum_{i=1}^{N_{e}} \psi_{i}(\vr,t) \int
  K(\vr,\vr')\psi_{i}^{*}(\vr',t)\varphi(\vr') \ud \vr'.
  \label{eqn:applyVX}
\end{equation}
This amounts to solving $N_{e}^2$ Poisson-like
problems with FFT, and the computational cost is $\Or(N_{g}\log
(N_{g})N_{e}^2)$, where the $N_{g}$ is the number of points in the
FFT grid. This cost is asymptotically comparable to other matrix
operations such as the QR factorization for orthogonalizing the
Kohn-Sham orbitals which scales as $\Or(N_{g}N_{e}^2)$, but the
$\log (N_g)$ prefactor is significantly larger. 

In order to propagate Eq.~\eqref{eqn:tddft} from an initial set of
orthonormal orbitals $\Psi(0)$, we may use for instance, 
the standard explicit 4th order Runge-Kutta scheme (S-RK4): 
\begin{equation}\label{eqn:srk4}
  \begin{split}
    k_1 &= -\I \Delta t H_n\Psi_n, \\
    \Psi_n^{(1)} &= \Psi_n + \frac{1}{2}k_1, 
    \quad H_n^{(1)} = H(t_{n+\frac{1}{2}},\Psi_n^{(1)}\Psi_n^{(1)*})\\ 
    k_2 &= -\I \Delta t H_n^{(1)}\Psi_n^{(1)}, \\
    \Psi_n^{(2)} &= \Psi_n + \frac{1}{2}k_2,
    \quad H_n^{(2)} = H(t_{n+\frac{1}{2}},\Psi_n^{(2)}\Psi_n^{(2)*})\\
    k_3 &= -\I \Delta t H_n^{(2)}\Psi_n^{(2)}, \\
    \Psi_n^{(3)} &= \Psi_n + k_3, 
    \quad H_n^{(3)} = H(t_{n+1},\Psi_n^{(3)}\Psi_n^{(3)*})\\
    k_4 &= -\I \Delta t H_n^{(3)}\Psi_n^{(3)}, \\
    \Psi_{n+1} &= \Psi_n + \frac{1}{6}(k_1 + 2k_2 + 2k_3 + k_4).
  \end{split}
\end{equation}
Here all the $H_n = H(t_n,P_n)$ is the Hamiltonian at step $t_n$, 
and $t_{n+\frac{1}{2}} = t_n + \frac{1}{2}\Delta t$, $t_{n+1} = t_n + \Delta t$. 
For each time step, the Hamiltonian operator needs to be applied $4$
times to each of the $N_{e}$ orbitals. After each update of the
orbitals, the Hamiltonian operator needs to be updated accordingly. The
maximal time step allowed by the RK4 integrator (and in general, all
explicit time integrators) is bounded by $c \norm{H}_{2}^{-1}$, where
$c$ is a scheme dependent constant and $\norm{H}_{2}$ is the spectral
radius of the Hamiltonian operator. In practice, this maximal time step
is often less than $1$ attosecond (as). Hence simulating rt-TDDFT for
$1$ fs would require more than $4000 N_{e}$ matrix-vector
multiplications involving the Hamiltonian operator (and hence the Fock
exchange operator). When the nuclei degrees of freedom are also
time-dependent such as in the case of the Ehrenfest dynamics, the
electron-nuclei potentials, such as the local and nonlocal components of
the pseudopotential, need also be updated more than $4000$ times per $1$
fs simulation.

\section{Parallel transport gauge}\label{sec:pt}

In order to accelerate rt-TDDFT calculations with hybrid functionals,
it is necessary to relax the constraint on the maximal time step that
can be utilized in the simulation. This can be achieved by the recently
developed parallel transport gauge
formalism~\cite{JiaAnWangEtAlPT,AnLinPT}, which we briefly
summarize below. 

First, note that Eq.~\eqref{eqn:tddft} can be equivalently written using
a set of transformed orbitals $\Phi(t)=\Psi(t) U(t)$, where the gauge
matrix $U(t)$ is a unitary matrix of size $N_{e}$.
An important property of the density matrix is that it is
gauge-invariant: $P(t)=\Psi(t)\Psi^{*}(t)=\Phi(t)\Phi^{*}(t)$.  Physical
observables such as energies and dipoles are defined
using the density matrix instead of the orbitals. The
density matrix and the derived physical observables can often oscillate
at a slower rate than the orbitals, and hence can be discretized with a
larger time step.  
Our goal is to optimize the gauge matrix, so that 
the transformed orbitals $\Phi(t)$ vary \textit{as slowly as
possible}, without altering the density matrix. 
This results in the following variational problem
\begin{equation}\label{eqn:minproblem}
  \min_{U(t)} \quad \norm{\dot{\Phi}}^2_{F},
  \ \text{s.t.} \ \Phi(t) = \Psi(t)U(t), U^{*}(t)U(t)=I_{N_{e}}.
\end{equation}
Here $\norm{\dot{\Phi}}^2_{F}:=\mathrm{Tr}[\dot{\Phi}^{*}\dot{\Phi}]$
measures the Frobenius norm of the time derivative of the transformed
orbitals.  The minimizer of~\eqref{eqn:minproblem}, in terms of $\Phi$, 
satisfies
\begin{equation}
  P\dot{\Phi}=0.
  \label{eqn:PTcondition}
\end{equation}
Eq.~\eqref{eqn:PTcondition} implicitly defines a gauge choice for each
$U(t)$, and this gauge is called the \emph{parallel transport gauge}.
The governing equation of each transformed orbital $\varphi_{i}$ can be
concisely written down as
\begin{equation}
  \I \partial_t \varphi_i = H\varphi_i - 
  \sum_{j=1}^{N_e}\varphi_j\left<\varphi_j|H|\varphi_i\right>, 
  \quad i = 1, \cdots, N_e,
  \label{eqn:pt-ket}
\end{equation}
or more concisely in the matrix form
\begin{equation}
  \I \partial_t \Phi = H\Phi -
  \Phi(\Phi^{*}H\Phi), \quad P(t) = \Phi(t)\Phi^{*}(t).
  \label{eqn:pt}
\end{equation}
The right hand side of Eq.~\eqref{eqn:pt} is analogous to the residual
vectors of an eigenvalue problem in the time-independent setup.  Hence
$\Phi(t)$ follows the dynamics driven by residual vectors and is
expected to vary slower than $\Psi(t)$.   We refer to the
dynamics~\eqref{eqn:pt} as the parallel transport (PT) dynamics, and
correspondingly Eq.~\eqref{eqn:tddft} in the standard Schr\"odinger
representation as the Schr\"odinger dynamics.

The PT dynamics only introduces one additional term, and hence can be
readily discretized with any propagator.  The standard explicit 4th
order Runge-Kutta scheme for the parallel transport dynamics (PT-RK4)
now becomes
\begin{equation}
    \begin{split}
        k_1 &= -\I \Delta t \{H_n\Phi_n-\Phi_n(\Phi_n^*H_n\Phi_n)\}, \\
        \Phi_n^{(1)} &= \Phi_n + \frac{1}{2}k_1, 
        \quad H_n^{(1)} = H(t_{n+\frac{1}{2}},\Phi_n^{(1)}\Phi_n^{(1)*})\\ 
        k_2 &= -\I \Delta t \{H_n^{(1)}\Phi_n^{(1)} - \Phi_n^{(1)}(\Phi_n^{(1)*}H_n^{(1)}\Phi_n^{(1)})\}, \\
        \Phi_n^{(2)} &= \Phi_n + \frac{1}{2}k_2,
        \quad H_n^{(2)} = H(t_{n+\frac{1}{2}},\Phi_n^{(2)}\Phi_n^{(2)*})\\
        k_3 &= -\I \Delta t \{H_n^{(2)}\Phi_n^{(2)} - \Phi_n^{(2)}(\Phi_n^{(2)*}H_n^{(2)}\Phi_n^{(2)})\}, \\
        \Phi_n^{(3)} &= \Phi_n + k_3, 
        \quad H_n^{(3)} = H(t_{n+1},\Phi_n^{(3)}\Phi_n^{(3)*})\\
        k_4 &= -\I \Delta t \{H_n^{(3)}\Phi_n^{(3)} - \Phi_n^{(3)}(\Phi_n^{(3)*}H_n^{(3)}\Phi_n^{(3)})\}, \\
        \Phi_{n+1} &= \Phi_n + \frac{1}{6}(k_1 + 2k_2 + 2k_3 + k_4).
    \end{split}
\end{equation}

We should note that the usage of the PT dynamics alone cannot enlarge
the time step if the spectral radius $\norm{H}_{2}$ is large. However,
when the dynamics becomes slower, this problem can be addressed by using
implicit time integrators.  For instance,
the Crank-Nicolson scheme for the Schr\"odinger dynamics (S-CN)
is
\begin{align}\label{eqn:scn}
    \left(I + \I \frac{\Delta t}{2} H_{n+1}\right)\Psi_{n+1}
    = \left(I -\I \frac{\Delta t}{2} H_n\right)\Psi_n,
\end{align}
while the Crank-Nicolson scheme for the parallel transport
dynamics (PT-CN) is
\begin{equation}\label{eqn:ptcnApp}
  \begin{split}
    &\Phi_{n+1} + \I \frac{\Delta t}{2} \left\{
    H_{n+1} 
    \Phi_{n+1} - \Phi_{n+1}\left(\Phi_{n+1}^{*}
    H_{n+1} \Phi_{n+1}\right)\right\} \\
    = &\Phi_{n} -\I \frac{\Delta t}{2} \left\{
    H_{n}
    \Phi_{n} - \Phi_{n}\left(\Phi_{n}^{*}
    H_{n} \Phi_{n}\right)\right\}.
  \end{split}
\end{equation}
When implicit time integrators are used, $\Phi_{n+1}$ needs to be solved 
self-consistently, which can be efficiently solved by mixing schemes
such as the Anderson method~\cite{Anderson1965}.  Numerical results
\REV{indicate} that the size of the time step for the PT-CN scheme can be
$10\sim 100$ as, and is significantly larger than that of standard explicit time
integrators.

We also remark that the computational complexity of standard rt-TDDFT
calculations may achieve $\Or(N_e^2)$
scaling~\cite{AlonsoAndradeEcheniqueEtAl2008,AndradeAlberdi-RodriguezStrubbeEtAl2012,Jornet-SomozaAlberdi-RodriguezMilneEtAl2015},
assuming 1) local and semi-local exchange-correlation functionals and
certain explicit time integrators are used, and 2) no orbital
re-orthogonalization step is needed throughout the simulation.  The PT
dynamics requires the evaluation of the term $\Phi(\Phi^*H \Phi)$ in
Eq.~\eqref{eqn:pt}, which scales cubically with respect to the system
size. We have demonstrated that the cross over point between the
quadratic and cubic scaling algorithms should occurs for systems
with thousands of atoms~\cite{JiaAnWangEtAlPT}.  In the current context
of hybrid functional rt-TDDFT calculations, both methods scales
cubically with respect to the system size due to the dominating cost
associated with the Fock exchange operator. Numerical results indicate
that the advantage of the PT formulation becomes even more evident in this
case.

%
%
%
%
%
%

\section{Adaptively compressed exchange operator formulation}\label{sec:pcdiis}

For hybrid functional rt-TDDFT calculations, the use of the parallel
transport gauge and implicit time integrators still require a relatively
large number of matrix-vector multiplication operations involving the
Fock exchange operator. For instance, when a relatively large time step
is used, the number of self-consistent iterations in each PT time step
may become $20\sim 40$. This gives room for further reduction of the
cost associated with the Fock exchange operator, using the recently
developed adaptively compressed exchange (ACE) operator
formulation~\cite{Lin2016ACE}. 

In ground state hybrid functional DFT calculations, every time when the
Fock exchange operator is applied to a set of orbitals, we store the
resulting vectors as
\begin{equation}
  W_{i}(\vr) = (V_{\X}[P]\varphi_{i})(\vr) \quad
  i=1,\ldots,N_{e}.
  \label{eqn:W}
\end{equation}
Here we assume that the Fock exchange operator is defined with respect
to the density matrix $P$, which is often also specified by the orbitals
$\{\varphi_{i}\}_{i=1}^{N_{e}}$. The vectors $\{W_{i}\}_{i=1}^{N_{e}}$
are then used to construct a \textit{surrogate} operator, or the
adaptively compressed exchange operator denoted by $\vace$. We require
that $\vace$ should satisfy the following consistency conditions
\begin{equation}
   (\vace \varphi_{i})(\vr) = W_{i}(\vr) \quad
   \mbox{and} \quad
   \vace(\vr,\vr') = \vace^{*}(\vr',\vr)
  \label{eqn:ACEcond}
\end{equation}
The conditions~\eqref{eqn:ACEcond} do not yet uniquely determine
$\vace$. However, the choice becomes unique if we require $\vace$ to be
strictly of rank $N_{e}$~\cite{LinLindseyACE}, and it can be computed as
follows. We first construct the overlap matrix
\begin{equation}
M_{ij} = \int \varphi^{*}_{i}(\vr) W_{j}(\vr) \ud \vr, \quad
i,j=1,\ldots,N_{e},
\end{equation}
which is Hermitian and negative definite.  We perform the Cholesky
factorization for $-M$, i.e. $M=-LL^{*}$, where $L$ is a lower
triangular matrix.  Then then the adaptively compressed exchange
operator is given by the following rank $N_{e}$ decomposition
\begin{equation}
  \vace(\vr,\vr') = -\sum_{k=1}^{N_{e}} \xi_{k}(\vr) \xi^{*}_{k}(\vr'),
  \label{eqn:ACE}
\end{equation}
where $\{\xi_{k}\}_{k=1}^{N_{e}}$ are called \textit{projection
vectors}, and are defined as
\begin{equation}
  \xi_{k}(\vr) = \sum_{i=1}^{N_{e}} W_{i}(\vr) (L^{-*})_{ik}.
  \label{eqn:Wtilde}
\end{equation}

The cost for applying $\vace$ to a number of vectors only involves
matrix-matrix multiplications up to size $N_{g}\times N_{e}$, which can
be efficiently carried out in the sequential or parallel settings.  The
preconstant of this step is also significantly smaller than that for
applying the Fock exchange operator. When self-consistency is reached,
$\vace$ agrees with the true Fock exchange operator when applied to the
occupied orbitals, thanks to the consistency
condition~\eqref{eqn:ACEcond}. We have also proved that for linear
eigenvalue problems, the ACE formulation can converge globally
from almost everywhere, with local convergence
rate favorable compared to standard iterative methods~\cite{LinLindseyACE}.

In order to utilize the ACE formulation in the context of rt-TDDFT
calculations, we note that the PT-CN scheme requires the solution of a
fixed point problem~\eqref{eqn:ptcnApp} for $\Phi_{n+1}$. Hence we may
artificially separate the fixed point problem into two iteration loops.
In the outer iteration, we apply the Fock exchange operator to
the parallel transport orbitals $\Phi_{n+1}$ only once, which gives rise
to $\vace$ defined by the procedure above.  Then in the inner iteration,
we perform a few inner iterations and only update the density-dependent
component of the Hamiltonian operator, while replacing the Fock exchange
operator by the same $\vace$ operator. This inner iteration step can
also be seen as a relatively inexpensive preconditioner for accelerating
the convergence of the self-consistent iteration.  Then we perform the
outer iteration until the density matrix (monitored by e.g. the Fock
exchange energy) converges. We summarize the resulting PT-CN-ACE
algorithm in  Alg.~\ref{alg:PTCNACE}, which propagates the
orbitals from the time step $t_{n}$ to $t_{n+1}$.

\begin{algorithm}[h]
  \caption{One step of propagation of the PT-CN-ACE method.}

  \begin{algorithmic}[1]

    \STATE Evaluate the right hand side of Eq.~\eqref{eqn:ptcnApp}, and
    choose an initial guess for $\Phi_{n+1}$ (the simplest choice being
    $\Phi_{n+1}=\Phi_{n}$).
    \WHILE {Fock exchange energy is not converged}

    \STATE Applying the Fock exchange operator to $\Phi_{n+1}$, and 
    construct $\vace$ in its low rank form. 

      \WHILE {electron density $\rho$ is not converged}
  
      \STATE Iteratively update $\Phi_{n+1}$ and $H_{n+1}$ using
      Eq.~\eqref{eqn:ptcnApp}, with $V_{X}$ replaced by $\vace$.
  
      \ENDWHILE
    \ENDWHILE

  \end{algorithmic}
  \label{alg:PTCNACE}
\end{algorithm}

\section{Numerical results} \label{sec:Result}


The S-RK4, PT-CN and PT-CN-ACE methods are implemented in the PWDFT package (based on
the plane wave discretization and the pseudopotential method),
which is an independent module of the massively parallel software package DGDFT (Discontinuous Galerkin 
Density Functional Theory)\cite{LinLuYingE2012,HuLinYang2015a}. PWDFT
performs parallelization primarily along the orbital direction, and can
scale up to several thousands of CPU cores for systems up to thousands of
atoms. We use the SG15 Optimized
Norm-Conserving Vanderbilt (ONCV)
pseudopotentials~\cite{Hamann2013,SchlipfGygi2015} and HSE06 functionals
\cite{HeydScuseriaErnzerhof2006} in all the following tests. \REV{We remark that
the SG15 pseudopotentials are obtained from all electron calculations using the PBE functional, which introduces certain amount of inconsistency in the hybrid functional calculation.}
The calculations are performed on the Edison supercomputer at National Energy Research 
Scientific Computing Center (NERSC). Each Edison node is equipped with two Intel Ivy Bridge 
sockets with 24 cores in total and 64 gigabyte (GB) of memory. Our code uses MPI only and the 
number of cores used is always equal to the number of MPI processes. 

In large scale hybrid functional TDDFT calculations, the application of
the Fock exchange operator dominates the total computational costs.
Hence we use the number of matrix-vector multiplications per orbital involving the Fock exchange
operator as a metric for the efficiency of a method, and this metric is
relatively independent of implementation.  In the case of PT-CN-ACE,
this number is equal to the number of times for which the ACE operator
need to be constructed. We also present the total wall clock time as
well as the breakdown of the computational time to properly take into
account contributions from other components, especially those
exclusive due to the usage of the PT-CN-ACE scheme.

\begin{figure}[h]
  \begin{center}
{\includegraphics[width=0.45\textwidth]{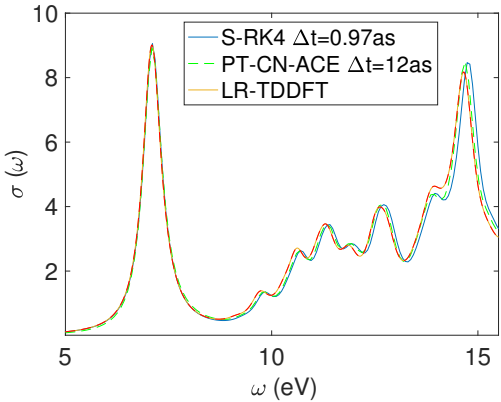}}
  \end{center}
  \caption{Partial absorption spectrum of benzene along the $x$ direction
  evaluated using the HSE06 functional.}
  \label{fig:benzene}
\end{figure}

We first demonstrate the accuracy of PT-CN-ACE by computing 
the absorption spectrum of the benzene molecule.
The size of the cubic supercell is $10.58\angstrom$ along each
direction, and the kinetic energy cutoff is set to
$20$ Hartree.  A $\delta$ kick is applied in the $x$ direction to calculate the partial 
absorption spectrum. The length of the rt-TDDFT calculation is 24 fs.
The size of the time step of PT-CN-ACE is set to be 12 as, and 
S-RK4 becomes unstable when the step size is bigger than 0.97 as. 
The absorption spectrum obtained by the S-RK4 and PT-CN-ACE methods are 
shown in Fig. \ref{fig:benzene} (b). We also provide benchmark results
obtained from the linear response time-dependent density functional theory (LR-TDDFT) 
calculation using the turboTDDFT module
\cite{MalciogluGebauerRoccaEtAl2011} from the Quantum 
ESPRESSO software package (QE) \cite{GiannozziBaroniBoniniEtAl2009}, which performs 3000 Lanczos steps 
along the $x$ direction to evaluate the component of the polarization tensor. 
Both QE and PWDFT use the same pseudopotential and kinetic
energy cutoff, and no empty state is used in calculating the spectrum in PWDFT.
A Lorentzian smearing of 0.27 eV is applied to all calculations. We find
that the shapes of the absorption spectrum calculated from three methods agree very well. 
The S-RK4 method requires $4$ Fock exchange operator calculations
per time step, while the PT-CN-ACE method only requires on average $3.2$
ACE operator constructions in each time step. Thus for this example, 
a total number of $98968$ and $6400$ Fock exchange operator applications per orbital are 
calculated for the S-RK4 and PT-CN-ACE method, respectively. This 
means that the PT-CN-ACE method is about 15 times faster than the S-RK4 method in
terms of the application of the Fock exchange operator.  
In the simulation, both PT-CN-ACE and S-RK4 use 
15 CPU cores, and the total wall clock time is 7.5 hours and 40.8 hours, respectively. 
The reduction of the speedup factor compared to the theoretical estimate
based on the number of Fock exchange operator applications is mainly due to 
the relatively small system size. Hence components such as the
evaluation of the Hartree potential, and the inner loop for solving the
fixed point problem in the PT-CN-ACE scheme still consume a relatively large portion of
the computational time.

\begin{figure}[h]
  \begin{center}
    \subfloat[Electric field along the $x$ direction]{\includegraphics[width=0.35\textwidth,angle=270]{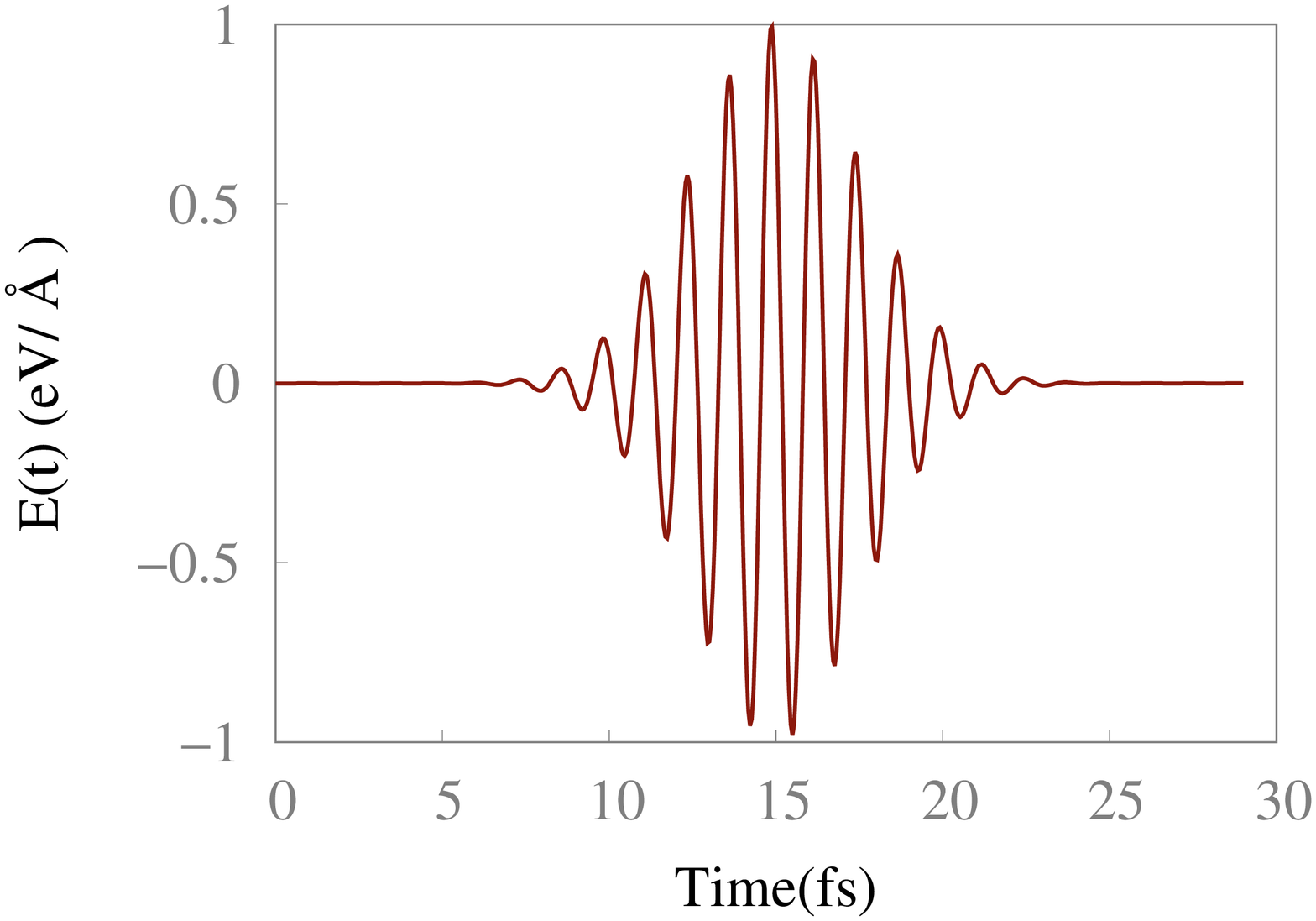}}
    \subfloat[Density of states at time $T$]{\includegraphics[width=0.35\textwidth,angle=270]{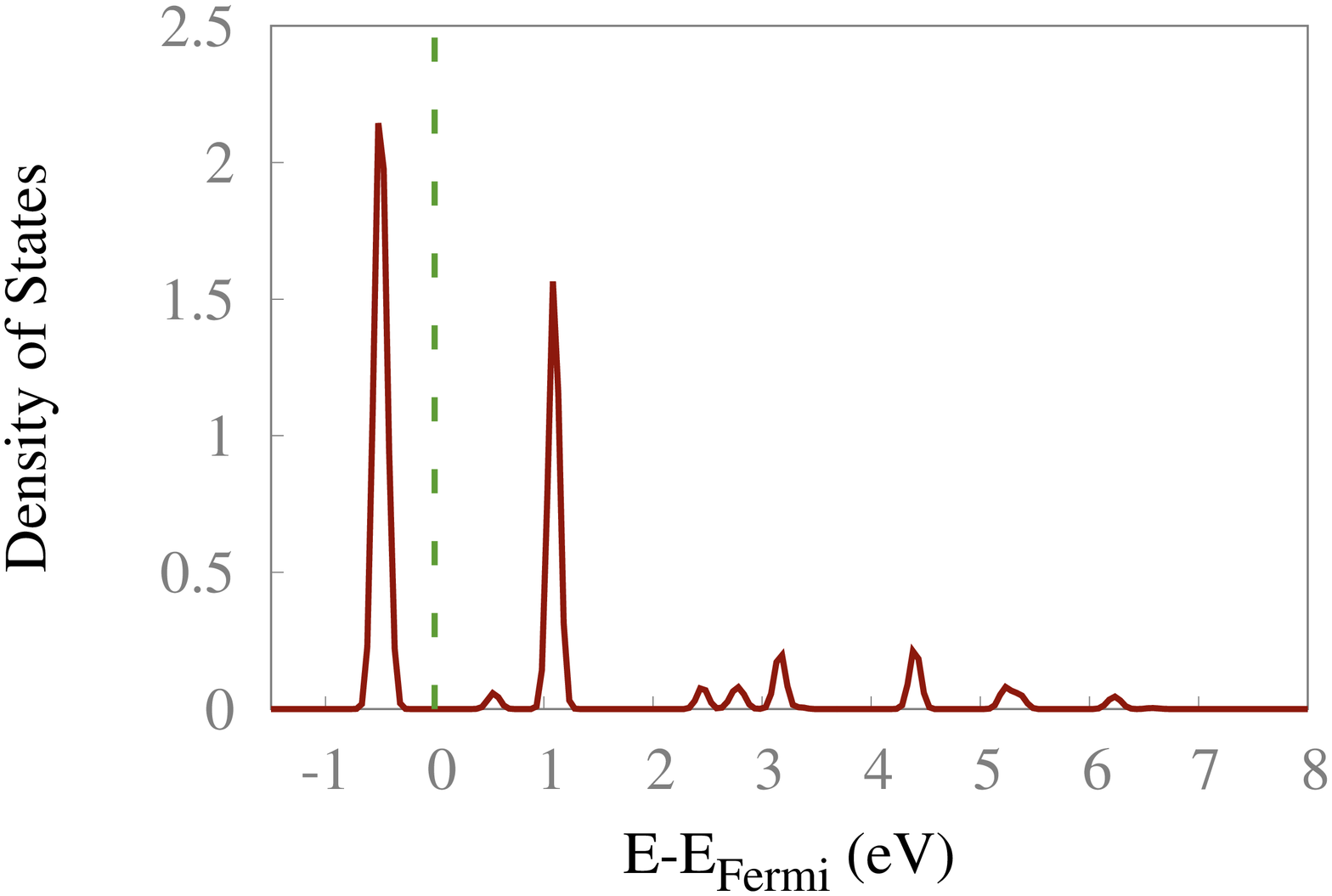}}
    \\
    \subfloat[Total energy]{\includegraphics[width=0.35\textwidth,angle=270]{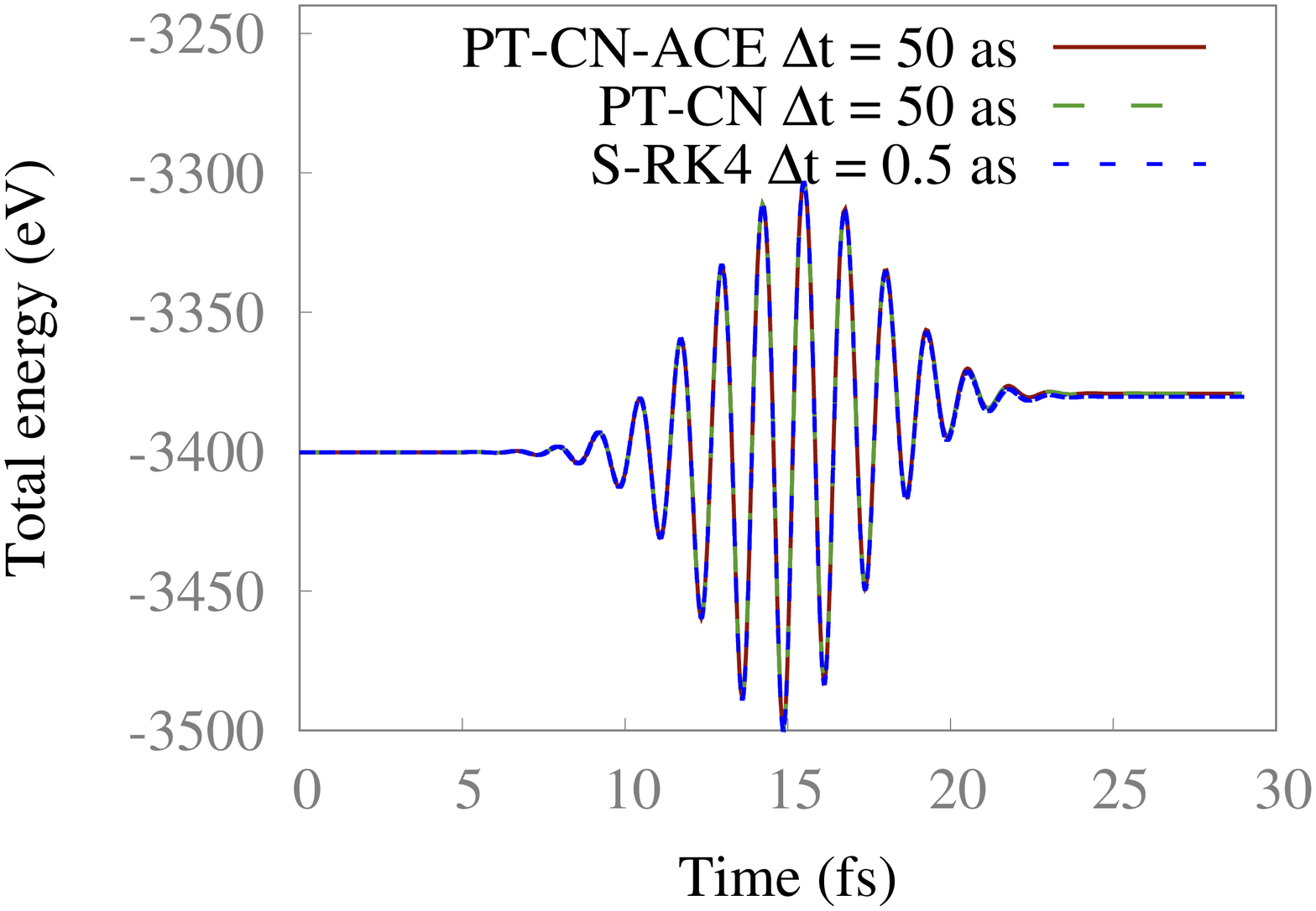}}
     \subfloat[Dipole moment along the $x$ direction]{\includegraphics[width=0.35\textwidth,angle=270]{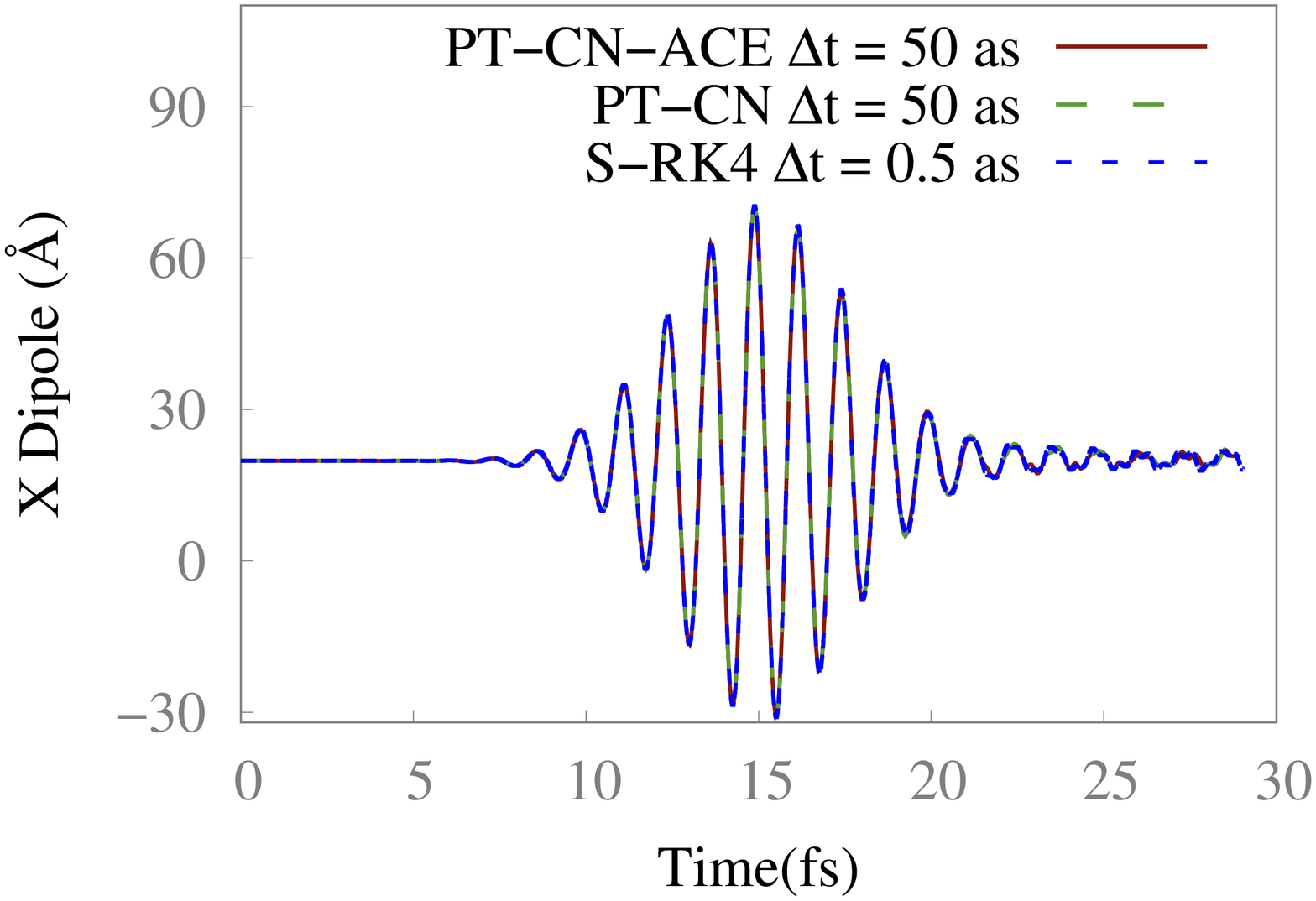}}
  \end{center}
  \caption{Electron dynamics of a 32 atom silicon system under a laser
  pulse. }
  \label{fig:ptcn-rk4}
\end{figure}

In the second example, we study the response of a silicon system
to an ultra-fast laser pulse.  The supercell consists of 32 atoms, and 
is constructed from $2 \times 2 \times 1$  unit cells \REV{sampled at the $\Gamma$ point}. Each simple
cubic unit cell has $8$ silicon atoms, and the lattice constant is
5.43 $\angstrom$.  The kinetic energy cutoff is set to 10 Hartree.  
The
laser is applied along the $x$ direction, and generates an electric
field of the form 
\begin{equation}
\textbf{E}(t) = \hat{\textbf{x}}E_{\text{max}}\exp \Big[-\frac{(t-t_0)^2}{2a^2}\Big]\sin [\omega(t-t_0)]{.}
\end{equation}
Here $a=2.55$ fs, $\hbar\omega=3.26 \mathrm{eV},t_{0}=15$ fs, 
which corresponds to a laser that peaks at $15$ fs with
its wavelength being $380$ nm. \REV{The laser intensity $E_{\text{max}}$ is 
0.0194 a.u. in the simulation.}
The profile of this external field is given in
Fig.~\ref{fig:ptcn-rk4} (a). The ground state band gap computed at $t=0$ is around
$1.0$ eV using the HSE06 functional with a supercell containing 64 atoms.
The total simulation length is $29$ fs.
In the PT-CN-ACE method, 
for the inner loop, the stopping criteria for the relative error of the electron
density is set to $10^{-7}$. The stopping criteria for the outer loop  is defined via 
the relative error of the Fock exchange energy, and is set to $10^{-8}$.
The stopping criteria for the PT-CN method is defined via the relative 
error of the electron density and is set to $10^{-6}$.
 \REV{We remark that although this is a periodic system and electron polarization should be evaluated using e.g. theories based on localized Wannier functions~\cite{King-SmithVanderbilt1993,MarzariMostofiYatesEtAl2012}, here for simplicity we just construct the laser field and measure the dipole moment by treating the silicon system as a large molecule. Such treatment is consistent among different choices of numerical measures, which is sufficient to demonstrate the efficiency and accuracy of the PT-CN-ACE algorithm.}



In order to demonstrate the electron excitation process,  we plot the
density of states at the end of the simulation (Fig. \ref{fig:ptcn-rk4}
(b), the green dotted line indicates the Fermi energy), defined as
\[
\rho(\varepsilon):=\sum_{j=1}^{N_{e}} \sum_{i=1}^{\infty}
|\langle\psi_i(T) | \varphi_j(T)\rangle|^2
\wt{\delta}(\varepsilon-\varepsilon_{i}(T)).
\]
Here $\varphi_{j}(T)$ is the $j$-th orbital obtained at the end of the TDDFT simulation at
time $T$, and $\varepsilon_{i}(T),\psi_i(T)$ are
the eigenvalues and wavefunctions corresponding to the Hamiltonian at
time $T$. $\wt{\delta}$ is a Dirac-$\delta$ function with a Gaussian
broadening of $0.05$ eV.  

Fig. \ref{fig:ptcn-rk4} (c), (d) show the total energy and the dipole moment
along the $x$ direction calculated with the PT-CN, PT-CN-ACE and S-RK4
methods.  The time steps of both PT-CN and PT-CN-ACE are set to 50 as, while
the time step of S-RK4 is set to 0.5 as due to stability reason.
Both the energy and the dipole moment agree very well among the results obtained
from these three methods, indicating that the use of the compressed
exchange operator in PT-CN-ACE does not lead to any loss of accuracy. 
The total energy and dipole moment obtained from S-RK4 and PT-CN-ACE are
nearly indistinguishable before $15$ fs, and becomes slightly different after 20 fs. 
Such error can be systematically reduced by using smaller time steps. 
Table \ref{tab:tddft_accuracy} shows that the error can be reduced to as 
small as $0.016$ meV/atom when the time step is reduced to 2.4 as for
PT-CN-ACE method. Nearly the same accuracy is also observed in the PT-CN method
when reducing the time step.
We also listed the speedup 
factors in terms of both the number of Fock exchange operator applications 
per orbital per time step (FOC)
and the total wall clock time (Wtime) in Table \ref{tab:tddft_accuracy}. 
The Wtime speedup is denoted in ``Speedup'' in Table \ref{tab:tddft_accuracy}. 
In comparison, the Fock exchange
operator application speedup, which is denoted as ``Speedup*'' in Table
\ref{tab:tddft_accuracy}, is calculated by counting the number of Fock 
exchange operator application for a given time period $\Delta$t. 
Note that the speedup factor obtained from the number of Fock exchange
operator applications is relatively bigger
than the wall clock time speedup, especially for PT-CN-ACE. 
This is mainly because the inner loop calculation in PT-CN-ACE still takes a big 
proportion of the computational time for this relatively small system. 
It is also why PT-CN can be faster than PT-CN-ACE in terms of the wall
clock time despite
of the fact that PT-CN requires more Fock exchange operator applications
per orbital. However, as the system size increases, the cost due to the Fock exchange 
operator becomes dominant, and we shall observe that PT-CN-ACE becomes
more advantageous below.


\begin{table}
  \centering
  \begin{tabular}{cc|cc|cc|cc}
  \hline Method &  $\Delta t$(as)& AEI(meV) &AED(meV) & FOC & Speedup* & Wtime(h) &Speedup\\\hline
         S-RK4     & 0.5    & 621.4156  &  /     & 4.0    & 1.0  &18.09 &1.0\\
         PT-CN     & 2.4    & 621.4062  &0.009   & 4.8    & 4    &6.0   &3.0\\
         PT-CN     & 5.1    & 621.4688  &0.053   & 5.3    &7.7   &3.65  &5\\
         PT-CN     & 12.1   & 623.4656  &2.05    & 5.9    &16.4  &1.45  &12.4\\ 
         PT-CN     & 25.0   & 628.8594  &7.44    & 10.8   &18.5  &1.08  &16.7\\
         PT-CN     & 50.0   & 657.6688  &36.25   & 21     &19    &1.12  &16.1\\
         PT-CN-ACE & 2.4    & 621.4313 & 0.016  & 2.32   & 8.3  &4.99   &3.7\\
         PT-CN-ACE & 5.1    & 621.4937 & 0.08   & 2.77   & 14.7 &3.11   &5.8 \\
         PT-CN-ACE & 12.1   & 623.5781 & 2.2    & 3.02   & 32.0 &1.6    &11.3\\
         PT-CN-ACE & 25.0   & 628.9531 & 7.5    & 3.78   & 52.9 &1.45   &12.5\\
         PT-CN-ACE & 50.0   & 657.725  & 36.3   & 5.8    & 69.0 &1.38 &13.1\\\hline
  \end{tabular}
  \caption{Accuracy and efficiency of PT-CN and PT-CN-ACE for the electron dynamics
  with the $380 nm$ laser compared to S-RK4. The accuracy is measured using
  the average energy increase per atom (AEI) after $29.0$ fs and the average 
  energy difference \REV{(AED)} per atom for PT-CN and PT-CN-ACE compared with S-RK4 method
  after $29.0$fs. The efficiency is measured 
  using the average number of Fock exchange operator applications per orbital in
  each time step (FOC) during the time interval from $0.0 $ fs to $29.0$ fs, 
  and the Fock exchange operator application speedup is denoted as ``Speedup*''. The total wall clock time (Wtime) 
  and the corresponding speedup factor are also listed.}\label{tab:tddft_accuracy}
\end{table}


\begin{figure}[h]
  \begin{center}
    \subfloat[Total wall clock time.]{\includegraphics[width=0.35\textwidth,angle=270]{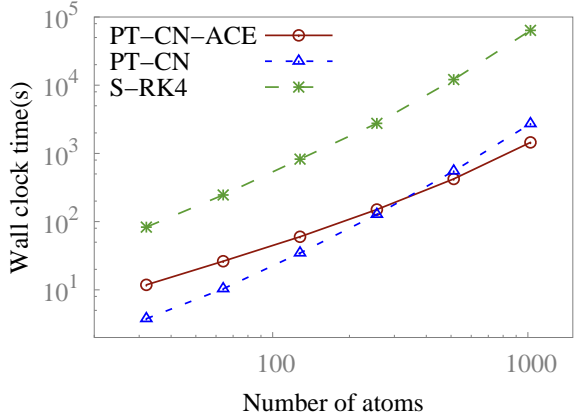}}\\
    \subfloat[Breakdown of the wall clock time, 1]{\includegraphics[width=0.35\textwidth,angle=270]{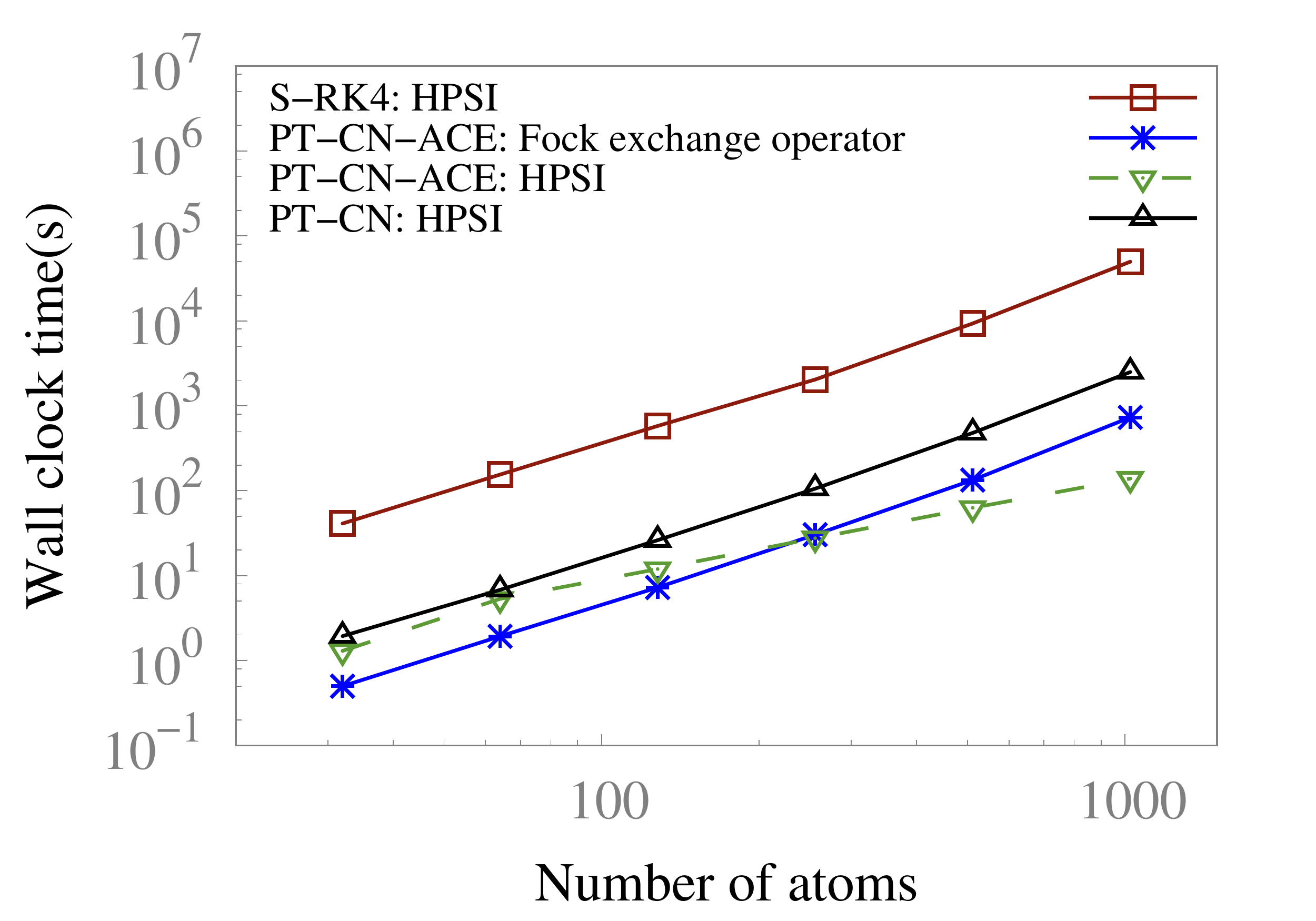}}
    \subfloat[Breakdown of the wall clock time, 2]{\includegraphics[width=0.35\textwidth,angle=270]{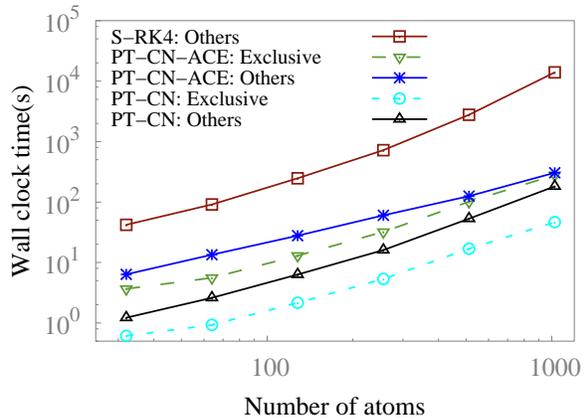}} \end{center}
  \caption{Wall clock time for simulating silicon systems from 32 atoms
  up to 1024 atoms for $\Delta t=50$ as. The number of CPU cores used
  are set to $2 \times N_{atom}$ in all tests. The systems are driven by
  the laser field shown at Fig. \ref{fig:ptcn-rk4}(a).}
  \label{fig:weakscaling}
\end{figure}

\begin{figure}[h]
  \begin{center}
    {\includegraphics[width=0.35\textwidth,angle=270]{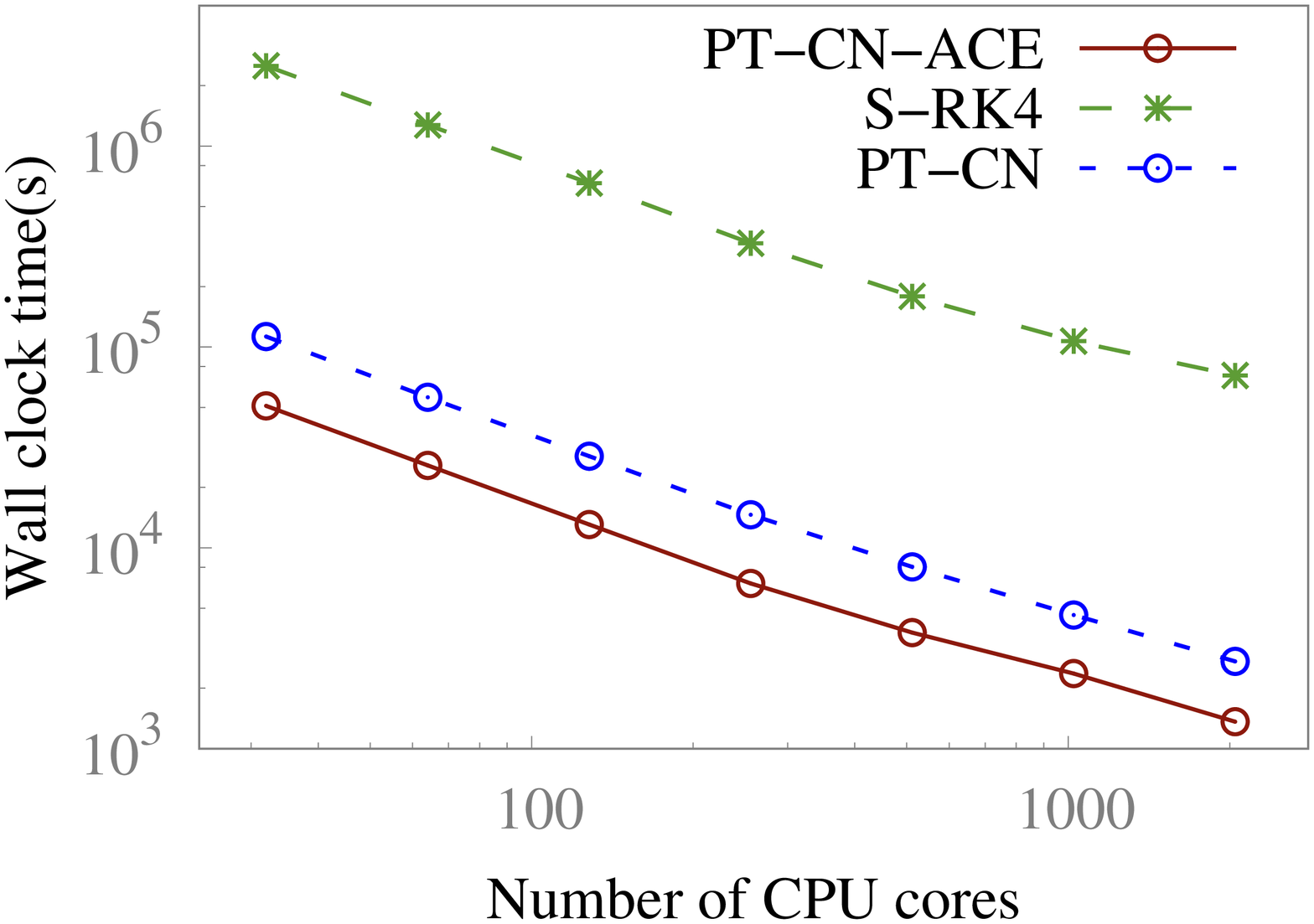}}
  \end{center}
  \caption{\REV{Total} wall clock time for 1024 atom silicon system from 32 up to 2048 CPU cores used in $\Delta t=50$ as. The system is driven by laser field shown at Fig. \ref{fig:ptcn-rk4}(a). }
  \label{fig:strongscaling}
\end{figure}

Next we systematically investigate the efficiency of the PT-CN and PT-CN-ACE
schemes by increasing the size of the supercell from $2 \times 2 \times
1$ to $8 \times 4 \times 4$ unit cells, and the set of
systems consists of $32$ to $1024$ silicon atoms.  
All other physical parameters remain the same as in the tests above.
We report the total wall clock time of the PT-CN, PT-CN-ACE and S-RK4
methods, as well as the breakdown into different components time of 
for a time period of $\Delta$t = 50 as.  We report the performance in terms of both the weak scaling
and the strong scaling.  The time step of PT-CN and PT-CN-ACE is set to be
50 as and the time step for S-RK4 is $0.5$ as. 
The average number of Fock exchange operator applications per orbital 
for PT-CN and PT-CN-ACE is $21$ and $5.75$, respectively. 
For the PT-CN-ACE method, the average number of inner iterations is
$24$.

The total wall clock time of PT-CN-ACE can be divided into 
four parts: ``ACE operator'', which stands for the time
used for applying the Fock exchange operator and constructing the ACE
operator implicitly; ``HPSI'', which represents the time for the
$H\psi$ calculation, with the application of the exchange operator
replaced by the application of the ACE operator via a matrix vector
multiplication operation; ``PT-CN-ACE: Exclusive'', 
which includes the time exclusively associated with the usage of the
parallel transport gauge, such as the orbital mixing and orbital
orthogonalization;
and ``Others'', which includes all other parts that are shared among the
three methods, such as the evaluation of
the Hartree potential and the total energy.  Similarly, the total wall clock
time of PT-CN is decomposed into three parts: ``HPSI'', ``PT-CN:
Exclusive'' and ``Others''.  The total wall clock time of S-RK4 is
divided into ``HPSI'' and ``Others''. Note that ``HPSI'' and the
evaluation of the total energy in PT-CN and
S-RK4 require the application of the true Fock exchange operator.

PWDFT is mainly parallelized along the orbital direction, i.e. 
the maximum number of cores is equal to the number of occupied orbitals. 
The application of the Fock exchange operator to the occupied orbitals
is implemented using the fast Fourier transformation(FFT).
In the ``HPSI'' component of the PT-CN-ACE method, the matrix matrix multiplication 
between the low rank operator $\vace$ and all the occupied orbitals is
performed to evaluate the Fock exchange term. 
We remark that certain components of PWDFT, such as the solution of the  Hartree potential, 
are currently carried out on a single core. This
is consistent with the choice of parallelization along the orbital 
direction, where each $H\psi$ (except the application of the Fock
exchange operator) is carried out on a single core. However,
as will be shown below, PT-CN-ACE and S-RK4 typically requires many more
Hartree potential evaluation compared to PT-CN. Hence PT-CN has some
advantage in terms of the wall clock time from this perspective.

Fig.~\ref{fig:weakscaling} (a) shows the total 
wall clock time with respect to the system size $N_{atom}$ for all 
three method. In these tests the number of CPU cores used
is always proportional to the number of atoms, i.e. $2\times N_{atom}$
(this is called ``weak scaling'').  
The speedup of PT-CN-ACE over S-RK4 is 7 times at 32 atoms, 
and increases to 46 times at 1024 atoms. On the other hand, 
the speedup of PT-CN over S-RK4 is between 22 and 23 times, which
is consistent with the Fock exchange operator applications speedup
as shown in Table \ref{tab:tddft_accuracy}. 
For small systems, PT-CN is the most efficient method. When the system
size increases beyond 256 atoms, PT-CN-ACE becomes the most efficient
one in terms of the wall clock time.
This cross-over can be explained by the breakdown of the total time 
shown in Fig.~\ref{fig:weakscaling} (b) (c). 
Since PT-CN-ACE method introduces a nested loop to reduce the number of Fock 
exchange operator applications, it also increases the number of inner
iterations. 
More specifically, in the tests above, the number of Fock exchange operator applications per orbital is 
$6$, but the number of inner iterations is 
$120$ in PT-CN-ACE. 
In comparison, PT-CN only requires $21$ 
inner iterations. Thus PT-CN is faster 
than PT-CN-ACE at small system size as shown in Fig. ~\ref{fig:weakscaling}(a). 
However, as system size becomes larger, the Fock exchange operator 
applications will dominate the cost, and PT-CN-ACE becomes
faster than PT-CN as shown in Fig. \ref{fig:weakscaling}(a). 


More specifically, Fig. \ref{fig:weakscaling}(b) shows that ``HPSI'' takes 49
to 78 percent of total wall clock time from 32 to 1024 atom system for
the S-RK4 method. 
For the PT-CN method, ``HPSI'' costs 51 percent of the time for the
system with 32 atoms, 
and this becomes 91 percent when the system size increases to 1024 atoms. 
For the PT-CN-ACE method, the cost involving the Fock exchange operator
is reduced to only 4 percent of the total time for the system with 32
atoms, and becomes 53 percent for the system with 1024 atoms.

Finally we report in Fig.~\ref{fig:strongscaling} the average wall clock time for carrying out
a simulation of $50$ as for the 1024 atom silicon system, with respect
to the increase of the number of computational cores (this is called ``strong scaling''). 
Compared to performance using 32 cores, the parallel efficiency of a 
single TDDFT step with 2048 cores reaches 54 percent, 58 percent and 64 percent for 
the S-RK4,  PT-CN-ACE and PT-CN methods, respectively. The reduction of
the parallel efficiency is mainly caused by our sequential 
implementation of certain components, such as the evaluation of the Hartree potential. 
The speedup of PT-CN-ACE method over S-RK4 is between 46 times and 50
times over the entire range.  Therefore in order to finish the
electron dynamics simulation above of $29$ fs,
it will take about 1 year using S-RK4, and this is reduced to around one
week using PT-CN-ACE.  Such a simulation is by all means still expensive,
but starts to become feasible to be routinely performed.

\section{Conclusion} \label{sec:conclusion}

In order to accelerate large scale hybrid functional rt-TDDFT
calculations, we have presented a method to combine two recently
developed ideas: parallel transport (PT) gauge and adaptively compressed
exchange (ACE) operator. The overall goal is to reduce the frequency for the application of the Fock exchange operator to orbitals, with systematically
controlled accuracy. We demonstrate that the resulting PT-CN-ACE scheme can indeed
reduce the number of Fock exchange operator applications per unit time
by one to two orders of magnitude compared to the standard explicit 4th
order Runge-Kutta time integrator, and thus enables hybrid functional
rt-TDDFT calculations for systems up to $1000$ atoms for the first time.

Compared to the PT-CN scheme, the extra reduction of the number of
applications of the Fock exchange operator requires more iterations in
the inner loop. This is consistent with the observation for ground state
hybrid functional calculations~\cite{HuLinYang2017a}. Hence in our implementation, PT-CN is in
fact faster than PT-CN-ACE in terms of wall clock time for small
systems. The precise cross-over point depends heavily on the cost for
solving the Poisson-like equation to apply the Fock exchange operator.
For instance, we expect that the PT-CN-ACE becomes advantageous at a
much earlier stage in real space rt-TDDFT software packages, where the
solution of a Poisson-like equation can be much more expensive than
that in a planewave based code. \REV{On the other hand, if the application of the Fock exchange operator can be accelerated using techniques such as localization~\cite{MarzariVanderbilt1997,DamleLinYing2015}, density fitting~\cite{LuYing2015,HuLinYang2017, DongHuLin2018}, or through the GPU architecture, we expect that the original PT-CN scheme will be more favorable.}
Further developments to reduce the
number of inner iterations without penalizing the number of Fock
exchange operator applications, such as via the usage of better
preconditioners, is also an interesting direction for future works.

\section{Acknowledgments}

This work was partially supported by the National Science Foundation
under Grant No. 1450372, No. DMS-1652330  (W. J. and L. L.), and by the
Department of Energy under Grant No. DE-SC0017867, No.
DE-AC02-05CH11231 (L. L.).  We thank the National Energy Research
Scientific Computing (NERSC) center and the Berkeley Research Computing
(BRC) program at the University of California, Berkeley for making
computational resources available.  We thank Dong An,
Zhanghui Chen and Lin-Wang Wang for helpful discussions.


\bibliography{ptref}
\newpage

\end{document}